\documentclass[12pt,preprint]{emulateapj}
\usepackage{amsmath, natbib, placeins,color, enumitem}
\usepackage{color}

\pdfoutput=1

\def\beq{\begin{equation}}
\def\eeq{\end{equation}}
\def\bey{\begin{eqnarray}}
\def\eey{\end{eqnarray}}
\newcommand{\gtorder}{\mathrel{\raise.3ex\hbox{$>$}\mkern-14mu
            \lower0.6ex\hbox{$\sim$}}}
\newcommand{\ltorder}{\mathrel{\raise.3ex\hbox{$<$}\mkern-14mu
            \lower0.6ex\hbox{$\sim$}}}

\begin{document}

\title{The Highest-Energy Cosmic Rays Cannot be Dominantly \\ Protons from Steady Sources}
\author{Ke Fang}
\affiliation{Department of Astronomy, University of Maryland, College Park, MD, 20742-2421, USA}
\affiliation{Joint Space-Science Institute, College Park, MD, 20742-2421}
\author{Kumiko Kotera}
\affiliation{Sorbonne Universit\'es, UPMC Univ. Paris 6 et CNRS, UMR 7095, \\ Institut d'Astrophysique de Paris, 98 bis bd Arago, 75014 Paris, France}
\affiliation{Laboratoire AIM-Paris-Saclay, CEA/DSM/IRFU, CNRS, Universite Paris Diderot,  F-91191 Gif-sur-Yvette, France}
\begin{abstract} The bulk of observed ultrahigh energy cosmic rays could be light or heavier elements, and originate from an either steady or transient population of sources.  This leaves us with four general categories of sources. Energetic requirements set a lower limit on single source luminosities, while the distribution of particle arrival directions in the sky sets a lower limit on the source number density. The latter constraint depends on the angular smearing in the skymap due to   the magnetic deflections  of  the charged particles during their propagation from the source to the Earth. We contrast these limits with the luminosity functions from surveys of existing luminous steady objects in the nearby universe,  and strongly constrain one of the four categories of source models, namely, steady proton sources. The possibility that cosmic rays with energy $> 8\times 10^{19}\,$eV are dominantly pure protons coming from steady sources is excluded at 95\% confidence level, under the safe assumption that protons experience less than $30^\circ$ magnetic deflection on flight.
\end{abstract}

\pacs{}
\maketitle

\section{Introduction}
The mystery of the origin of ultrahigh-energy cosmic rays (UHECR) remains unsolved \citep{KO11}. Observationally, one major obstacle stems from the lack of convergence on the measured chemical composition at the highest energies. The two leading UHECR observatories report puzzling results: while the Pierre Auger Observatory in the southern hemisphere finds medium to heavy composition above 50 EeV (1 EeV = $10^{18}$ eV) \citep{Aab:2015bza}, the Telescope Array in the northern hemisphere reports that the composition of particles above EeV is consistent with pure proton \citep{Fukushima:2015bza}. 

From a theoretical point of view, many promising UHECR candidate sources have been proposed in the literature. The extreme energy of UHECRs sets a lower limit on the bolometric luminosity of their accelerators \citep{1995PhRvL..75..386W}.
But none of the objects passing that cut has been tested conclusively yet -- positively or negatively.   Based on their decay timescale,
 they can be grouped into two categories: steady and transient sources. 
 A source can be categorized as steady if its  emission timescale is longer than the spread in the arrival time of their UHECRs \citep{2009JCAP...08..026W, 2012ApJ...748....9T}.   In this case, the arrival directions of UHECRs can directly trace and  constrain  the sky distribution of their sources, in conjunction with other neutral messengers like photons, neutrinos and gravitational waves. Such a spread is caused by magnetic deflections of charged cosmic rays in Galactic and intergalactic media, which can be quantified as $\delta t\approx  10^5\,(l/100 \,{\rm Mpc})\left(  \alpha/2^\circ\right)^2\,{\rm yrs}$ \citep{KL08b},  for a propagation distance $l$ and a total deflection angle $\alpha$.   We note that the definition of steadiness is relative and dependent on $l$ and $\alpha$.  For protons, $\alpha$ is typically a few degrees (as will be discussed at the end of Section~\ref{sec:LF}).   Thus $\delta t$ ranges from a few tens of years for a Galactic or local source, to $\gtrsim 10^4$\, years for a source at  the GZK horizon (\citealt{G66,ZK66}, also see Section~\ref{sec:LF}). 
 Examples of potential steady sources include radio-loud active galactic nuclei (AGN),  quasar remnants, and cluster accretion shocks.  
Examples of  transient candidates include gamma-ray bursts, fast-rotating neutron stars, and giant AGN flares   (see  \citealp{KO11} and references therein). 

The sources of UHECRs can thus be grouped into four major categories:   steady proton, steady heavy nuclei, transient proton and transient heavy nuclei sources. This letter examines whether the measured highest-energy cosmic rays could be protons from steady sources, by comparing the luminosity functions from surveys of luminous steady objects in the nearby universe with the required levels of number density and luminosity of UHE proton sources. We exclude at 95\% confidence level (C. L.) the possibility that the observed  highest-energy events could be dominated by pure protons from steady sources, under robust assumptions on the magnetic deflections experienced by particles during their propagation.   The remaining choices for sources of UHECRs are thus either steady accelerators of heavy elements or transients. 

\section{Luminosity and number density constraints}\label{sec:LF}
Above 60 EeV, observable sources must lie within the so-called GZK horizon, due to energy losses via interactions between extragalactic cosmic rays and the cosmic background radiation \citep{G66,ZK66}. The horizon is of order $100-200\,$Mpc at 60 EeV, implying that particles at the highest energies are produced in an anisotropic Universe. However, the arrival directions of observed UHECRs do not display any significant clustering other than in the Telescope Array hotspot region \citep{Abbasi:2014lda} and the  Centaurus A region \citep{Abreu2010314}. The lack of strong anisotropy sets lower bounds on the number density of sources, depending on the assumed magnetic deflection of particles \citep{AugerBound}. 

Using the level of clustering in the sky of selected events with energy thresholds of 60, 70, and 80 EeV, the Auger Collaboration  derived lower bounds on the source number density of $n_s\sim(0.06-5)\times10^{-4}\,\rm Mpc^{-3}$ at 95\% C. L., if sources are uniformly distributed and equally luminous \citep{AugerBound}. Similar 95\% C. L. bounds of $\sim(0.2-7)\times10^{-4}\,\rm Mpc^{-3}$ were derived for sources following local matter distribution. These bounds are subject to a factor of 3 uncertainty due to systematic errors on the  cosmic-ray  energy calibration. The ranges quoted for the bounds correspond to different assumed magnetic deflections. The most (least) stringent bound was obtained for the angular scale $\alpha=3^\circ$ ($\alpha=30^\circ$).

The lower bounds on the magnetic luminosity of an UHECR source haven been studied in  e.g., \cite{1995PhRvL..75..386W,LW09}.  The magnetic power contained in the magnetized plasma of an astrophysical source  can be written as $L_B = \beta c\,u_B \,4\pi\, R_{\rm acc}^2$, where $\beta c$ is the speed of the magnetic flow,  $u_B = B^2/4\pi$ is the magnetic energy density, and $R_{\rm acc}$ is the size of the acceleration region.  The  potential drop generated by the moving plasma is given by $V = \beta  B R_{\rm acc}/\Gamma$, where $\Gamma=\left(1-\beta^2\right)^{-1/2}$ is the Lorentz factor of the relativistic flow,   and $R_{\rm acc}/\Gamma$ is the effective size of the acceleration region, considering that the available time in the comoving frame is shortened by $\Gamma$. A charged particle passes the acceleration region would gain  energy $E_{\rm CR} \leq  Z e\beta  B R_{\rm acc}/\Gamma$. This sets a general lower bound to the magnetic luminosity  
\begin{equation}\label{eqn:LB}
L_B\geq\frac{\Gamma^2c}{\beta}\left(\frac{E}{Ze}\right)^2 >2\times 10^{45}\left(\frac{E}{Z\,80\,\rm EeV}\right)^2\,\rm erg/s\ .
\end{equation}
Note that equation~(\ref{eqn:LB}) is a universal argument  regardless of the acceleration mechanism or geometry of the acceleration region \citep{LW09}.  The term $\Gamma^2/\beta$ is larger than unity for a non-relativistic or sub-relativistic  source, and is comparable to the beaming factor  $\sim \Gamma^{-2} $   in  case of a relativistic outflow, hence the second inequality. 
The equipartition hypothesis suggests an equality of the energies in relativistic particles and magnetic field \citep{2011hea..book.....L}.  For most astrophysical objects, especially those not dominated by non-thermal emissions, $L_B$ is not expected to exceed  the bolometric luminosity, $L_{\rm bol}$. This can be violated if the source is dominantly powered by Poynting flux, and the majority of the magnetic energy is not dissipated into radiation.

\begin{figure}
\includegraphics[scale=0.45]{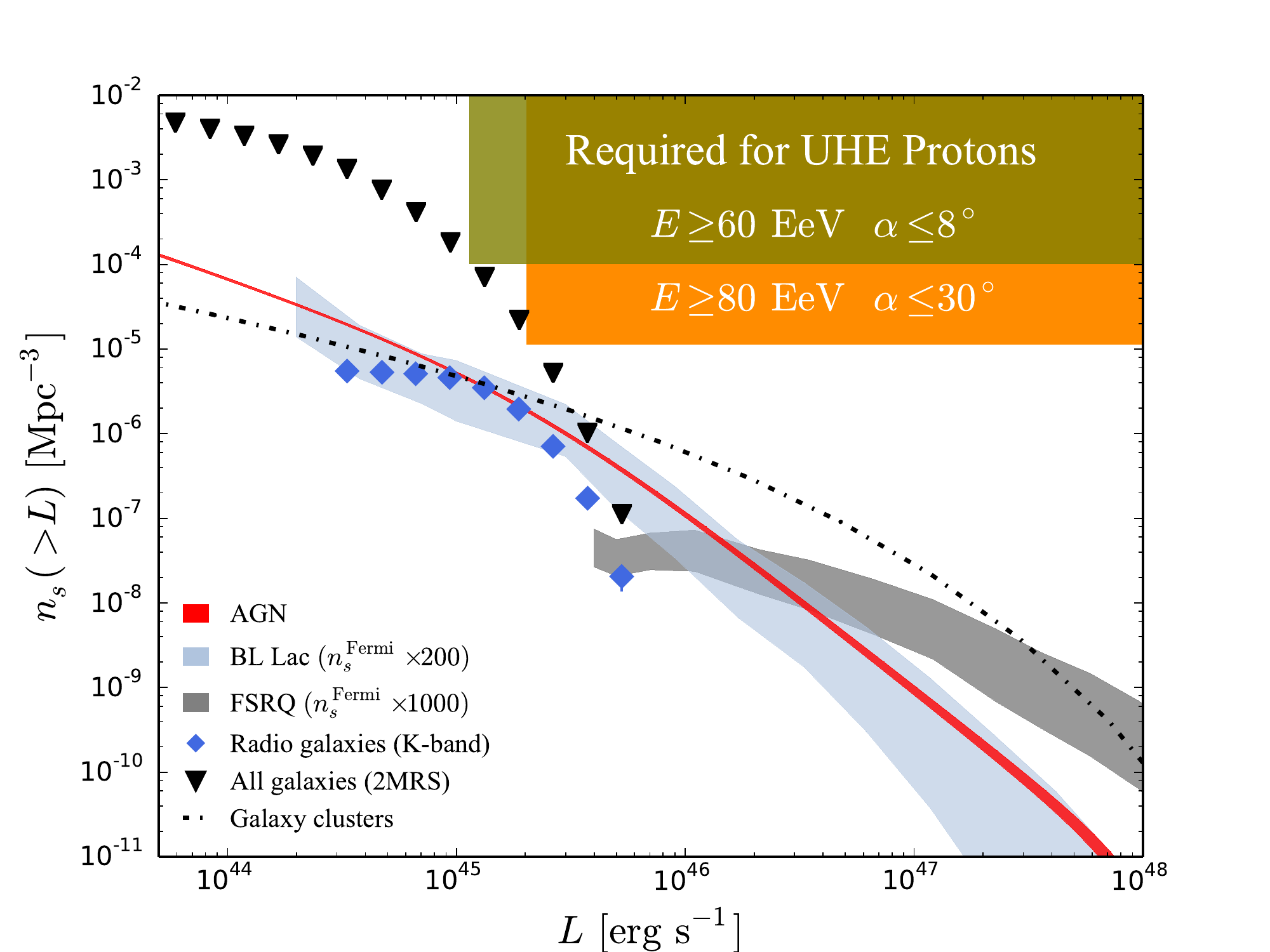}
\caption{\label{fig:nL} {Source number density and luminosity required by UHECR observations, assuming that cosmic rays are protons from steady sources (shaded in orange and  green boxes), compared to the cumulative luminosity functions of bright galaxies in the nearby universe (black triangle markers) \citep{2012A&A...544A..18V, 2014A&A...567A..81R}, AGN from X-ray surveys (red band)  \citep{2014ApJ...786..104U}, radio galaxies (blue diamond makers) \citep{2012A&A...544A..18V},   BL Lac objects (light blue shaded band)  \citep{Ajello:2013lka}  and flat spectrum radio quasars (grey shaded band)  \citep{2012ApJ...751..108A} observed by the Fermi Telescope \citep{2011ApJ...743..171A}, and galaxy clusters (dash-dotted black line).   The reported luminosities are assumed to be representative of the magnetic luminosity (see text for discussion). The lower bound on the source number density at 95\% C. L. is obtained assuming a deflection angle of UHECRs  $\alpha \le 8^\circ$ (green), and $\le 30^\circ$ (orange). The lower limit on the source luminosity corresponds to the magnetic luminosity of a source that can accelerate particles to $\ge 60$ (green) and $\ge 80$ (orange) EeV (Eq. 1). }
}
\end{figure}

In light of these requirements, we perform a census of the luminosity functions (LF) of known bright sources in relevant wavebands and report them in Figure~\ref{fig:nL}.
For a conservative comparison and to take into account the uncertainty arising from the conversion of the luminosity in observed bands to $L_B$ for some sources, we consider a cumulative luminosity function. In the plot, the reported luminosities are assumed to be representative of the magnetic luminosity, under the equipartition hypothesis. The issues related to such an assumption, and to the luminosity conversions are discussed below. 

\section{Comparison with luminosity functions from surveys}
The 2MASS Redshift Survey (2MRS) \citep{2MRS}   maps the all-sky three-dimensional distribution of galaxies in the nearby universe in the near infrared band. It is the best survey that describes the matter distribution in the nearby universe, and it provides a general measurement of the distribution of bright sources regardless of specific source types. 
 The K-band all-galaxy LF \citep{2012A&A...544A..18V,2014A&A...567A..81R}  derived from the catalog 
 is indicated as black triangle markers in Figure~\ref{fig:nL} (using a cosmological constant $h=0.678$  \citep{2015arXiv150201589P}).

A complete X-ray LF of AGN was provided by \cite{2014ApJ...786..104U}, utilizing  combined samples from surveys performed with Swift/BAT, MAXI, ASCA, XMM-Newton, Chandra and ROSAT.  The  bolometric luminosity of AGNs can be derived from the X-ray luminosity by a luminosity-dependent bolometric correction    \citep{2007ApJ...654..731H}. The bolometric LF for AGNs in the local universe is shown as a red  band in Fig~\ref{fig:nL},  taking   $\Omega_m=0.308$  \citep{2015arXiv150201589P} and assuming a flat Universe.  
An alternative bolometric correction obtained from    simultaneous optical-to-X-ray spectral energy distributions of  hard X-ray-selected local AGNs \citep{2009MNRAS.399.1553V}   leads to a similar LF.

Among  the galaxy population, radio-loud active galaxies have been suggested to satisfy necessary preconditions to accelerate and confine UHECRs (e.g. \citealp{1984ARA&A..22..425H}). An all-sky catalog of extragalactic radio sources of the local universe is provided by \cite{2012A&A...544A..18V}. The catalog was obtained by matching  radio-emitting galaxies from existing radio catalogs, including surveys from NVSS  \citep{1998AJ....115.1693C} at 1.4 GHz and SUMSS  \citep{1999AJ....117.1578B, 2003MNRAS.342.1117M} at 843 MHz,  with galaxies from the 2MRS \citep{2MRS}. The K-band LF for powerful radio galaxies in this catalog ($L>10^{24}\,\rm W\, Hz^{-1}$) is shown by blue diamond markers in Figure~\ref{fig:nL}.  Radio galaxies constitute about 20\% of the total galaxy population above $\sim10^{45}\,\rm erg\,s^{-1}$   \citep{2012A&A...544A..18V}. 
We caution that the relationship between the electromagnetic energy of radio-galaxies and radio/near infrared observations is unknown. Allowing for modeling uncertainties, the LF could be significantly shifted horizontally. Although the standard scenarios favor $L_B<L_{\rm bol}$ (e.g., \citealp{Merloni07}), Poynting-flux dominated models also exist (e.g., \citealp{Nakamura08}). An order of magnitude shift with $L_B\sim 10\,L_{\rm bol}$ would however not change our conclusions.

Blazars represent an extreme subclass of the radio-loud AGN, with a relativistic jet pointing along the line of sight of the Earth. Due to their extremely powerful jets, the $\gamma$-ray luminosity of blazars could be comparable or even higher than the total luminosity in other bands \citep{1996ApJ...463..444S}.  
The Fermi Telescope has provided the largest sample of  blazars to date in $\gamma$-rays \citep{2011ApJ...743..171A}.    LFs of  flat spectrum radio quasars (FSRQs) and   BL Lacertae (BL Lac) objects  have been derived using the  $\gamma$-ray-selected blazars  \citep{2012ApJ...751..108A, Ajello:2013lka}. Due to relativistic beaming, only a small fraction of these objects can be observed from the Earth. 
FSRQs are most seen within $5^\circ$ of the jet axis with a peak at $2^\circ$, while BL Lac objects are most seen within $10^\circ$ of the jet axis with a peak around $5^\circ$  \citep{2012ApJ...751..108A, Ajello:2013lka}. Assuming a random distribution for the angle between a jet axis and the line of sight ($P(\theta)=\sin(\theta)$ with $\theta$ being the viewing angle), the observed FSRQ and BL Lac samples represent   $\sim0.1\%$   \citep{2012ApJ...751..108A} and $\sim0.5\%$ of the parent population. 
 As opposed to photons, UHECRs from an off-aligned AGN  could still be deflected in the magnetic field later during their propagation and be reorientated into the direction of the Earth. Therefore a fraction of the off-aligned AGNs   could also contribute to UHECRs, depending on the level of deflection.  
In Figure~\ref{fig:nL} we take the Fermi LFs of FSRQ and BL Lac objects de-evolved at redshift 0   and multiple their number density by $1000$ and $200$   respectively to estimate the LF of the entire blazar population. 
Note that due to the large uncertainties in the LFs from  \cite{2012ApJ...751..108A} and \cite{Ajello:2013lka} we choose to show $n(L)$ instead of $n(>L)$ in Figure~\ref{fig:nL}. The difference is however negligible.  
The amplified $\gamma$-ray LF is consistent with the LF of radio galaxies, and comparable to the  bolometric LF of the  entire AGN population within uncertainties, implying a common population, as suggested by the unification scenario \citep{1995PASP..107..803U}. 
Because UHE protons are not expected to deflect  completely isotropically
and only a fraction of the Fermi blazars are possibly hosting hadronic processes,  the amplified LF represents an upper limit to   what  can be reached by the sources of UHECRs. 
It is again unclear whether blazar outflows are dominated by kinetic or Poynting-flux power, and whether the magnetic luminosity exceeds the bolometric (see  \citealp{2008MNRAS.385..283C} and refs. therein), but our results remain valid within an order of magnitude shift of the LF.

Another type of steady candidate is galaxy clusters  \citep{1984ARA&A..22..425H}. The upper limit on the energy of the turbulent magnetic field   is determined by the rate of accretion of matter onto  the cluster. 
The total accretion energy reads   $L_{\rm acc}=f_b\,GM\dot{M}/r_{\rm vir}$    \citep{Murase:2008yt, FO16}, 
 where  $f_b=0.13\,(M/10^{14}M_\odot)^{0.16}$ is the average baryon fraction of galaxy clusters \citep{Gonzalez:2013awy}, and  $r_{\rm vir}$ is the virial radius of the cluster. The mass accretion rate is confined by observations as $\langle \dot{M}\rangle (z=0)=42\,(M/10^{12}\,M_\odot)^{1.13}\,\rm M_\odot\,yr^{-1}$ \citep{McBride:2009ih}. The corresponding number density is obtained by integrating the cluster halo mass function  \citep{sheth_tormen_mass_function}  at redshift 0.

In Figure 1 we contrast the above LFs with the  region where sources must belong in order to produce the highest-energy cosmic ray protons. 
The orange box, corresponding to steady proton sources above 80 EeV,  comes directly from the least stringent bound of \cite{AugerBound}  (for deflection angles of $30^\circ$) and   is disjoint from  all the observed LFs. 
  The green box in Figure~\ref{fig:nL} corresponds to sources of particles with energy above 60 EeV, that experience magnetic deflections of $\alpha\le 8^\circ$. Its lower limit on $n_s$ was chosen by scanning over the range of deflection angles explored by \cite{AugerBound} and selecting the lowest density allowed at 95\% C.L. that does not overlap with the 2MRS luminosity function. Note that, as we discuss below, larger deflection angles would lead to an overlap of the allowed region with a 2MRS population of galaxies that are not expected to be sources of UHECRs. Deflections of $\alpha \le 22^\circ$ would be allowed if matching the lower limit to the AGN LF.

The level of UHECR deflection depends mainly on the strength and configuration of the extragalactic magnetic fields, which are highly uncertain. The observational bounds that have been placed on the global field strength $B$ and coherence length $\lambda$ cover a wide range: $B\lambda^{1/2}\in [  10^{-19}-10^{-8}]\,$G\,Mpc$^{1/2}$ \citep{Ryu98}.  Heavy numerical simulations of the cosmic magnetic fields and of particle propagation lead to discrepant results (e.g., \citealp{Das08}), and often fail at modeling adequately the diffusion of particles due to computational limitations. Semi-analytical models can be used however to infer that a standard set of Galactic and extragalactic magnetic fields should lead to proton deflections of order a few degrees above GZK energies \citep{WM96,KL08b}. 

Quantitatively, the deflection that particles of energy $E$ and charge $Z$ experience due to  a homogeneous intergalactic magnetic field {with strength B and coherence length $\lambda$ over a distance $d$ reads} $\alpha\sim 6.3^\circ\,Z(E/60\,{\rm EeV})^{-1}(d/100\,{\rm Mpc})^{1/2}(\lambda/{\rm Mpc})^{1/2}(B/{\rm nG})$ \citep{WM96}. If magnetic inhomogeneities are taken into account for a more realistic modeling, the deflection can be expressed as \citep{KL08b}:
\begin{eqnarray}
\alpha& \sim &2^\circ\,Z\,\left(\frac{E}{60\,\rm {EeV}}\right)^{-1}\left(\frac{\tau}{3}\right)^{1/2}\left(\frac{r_{\rm i}}{2{\,\rm Mpc}}\right)^{1/2}\times \nonumber\\ 
&&\left(\frac{B_{\rm i}}{10 \,{\rm nG}}\right)\left(\frac{\lambda_{\rm i}}{0.1\,\rm{Mpc}}\right)^{1/2},
\end{eqnarray}
where magnetized regions (such as filaments, radio ghosts, clusters of galaxies), are characterized by their typical size $r_i$, magnetic field coherence length $\lambda_i$ and strength $B_i$. Trans-GZK particles propagating in the intergalactic medium  typically encounter a  number $\tau\sim 3$ of such regions \citep{KL08b}. 
The propagation in the Galactic magnetic field results in an additional deflection of $\alpha_{\rm Gal}$, the quadratic sum of the turbulent ($\alpha_{\rm turb}$) and regular ($\alpha_{\rm reg}$) components. Numerically, $\alpha_{\rm turb} \sim 0.5^\circ\,Z(E/60\,{\rm EeV})^{-1}(H_{\rm Gal}/2\,{\rm kpc})^{1/2}(\lambda_{\rm Gal}/50\,{\rm pc})^{1/2}\times(B_{\rm Gal, turb}/3\,\mu{\rm G})$,
where $B_{\rm Gal}$, $\lambda_{\rm Gal,turb}$ and $H_{\rm Gal}$ are the magnitude, coherence length, and height of the turbulent component of the Galactic magnetic field, and $\alpha_{\rm reg} \sim 3.5^\circ\,Z(E/60\,{\rm EeV})^{-1}(L_{\rm Gal}/2\,{\rm kpc})(B_{\rm Gal,reg}/2\,\mu{\rm G})$, for a field coherent over lengthscale $L_{\rm Gal}$ and of strength $B_{\rm Gal,reg}$  \citep{KST07}.
The above values for the Galactic field are only indicative, and larger deflections up to $\sim10^\circ$
could be obtained for other configurations of the magnetic field within the observationally constrained range  \citep{Haverkorn15}. Overall deflections of $\le 8^\circ$ for protons, as quoted above, can thus be viewed as highly reasonable, and $\le 30^\circ$ is extremely robust.


\section{Conclusion and discussion}
The requirements for the production of cosmic ray protons at the highest energies in terms of source luminosity and number density are tight enough to exclude steady candidate sources. Because the observational constraints on the source number density depend on the particle deflection angles, this exclusion relies on the comfortable assumption that protons are deflected of $\le 30^\circ$ above $80\,$EeV. 
Our 95\% C. L. on the exclusion directly results from the allowed $n_s$ region (for a given deflection angle) quoted by  \cite{AugerBound}.  We stress that because our exclusion statements are directly related to the deflection angle, they are not subject to the large uncertainties and subtle details of magnetic field measurements. The luminosity limit is a theoretical pre-requisite and does not have a C. L. attached.

The K-band LF from the 2MRS survey was used as an estimation to the bolometric LF of normal galaxies. If $L_{\rm bol}$  were significantly higher than $L_{\rm K-band}$ (which is possible for some subset of the galaxy population, especially those with strong star formation activities),  our  conclusion would still be valid with reasonably  smaller  deflection angles. 
Notice that the population of sources dominating a K-band survey sample above $L\sim 10^{45}\,$erg/s
are known to be mostly passively-evolving, early-type galaxies, located at the center of galaxy clusters (e.g., \citealp{Bonne15}). In spite of their energy budget, their quiescence and the absence of associated high-energy emission makes them very difficult to reconcile with the production of UHECRs. 

The exclusion of steady proton sources could appear even stronger if the boxes were compared to the LFs of radio galaxies and blazars, from which high-energy emission has been detected, and the constraints on the deflection angles would be much relaxed. 
We recall  that the bounds on the source number density reported in Figure~\ref{fig:nL} are the conservative values quoted by \cite{AugerBound} for an uniform source distribution.  The constraints on $n_s$ would be   $\sim 3 - 10$ times better if considering the inhomogeneous distribution of the local structures.
This effect is however absorbed by the uncertainties due to the energy calibration of the observed cosmic rays, that leads to uncertainties of the same order \citep{AugerBound}.  The major unknown remains however the relationship between the bolometric and the magnetic luminosities of the source. If the latter dominates significantly (and if this property is shared by a large fraction of sources within a given population), the LFs would have to be shifted to the right for an accurate comparison with the proton steady source box.

Our conclusions concern the dominant sources of UHECRs and does not exclude the existence of a steady source contributing to the observed spectrum at a minor rate, that would not affect the anisotropy analysis of \cite{AugerBound}.

If one alleviates the primary proton assumption, lower luminosity sources would pass the cut (Eq.~\ref{eqn:LB}), enlarging the allowed parameter space to the left in Fig.~\ref{fig:nL}. As for the source number density, \cite{AugerBound} comments that, although their analysis was performed using protons, the propagation of iron nuclei leads to similar results (for given deflection angles) because the energy loss rates due to photo-disintegration of iron nuclei on cosmic backgrounds are comparable to those of protons. The deflections being stronger for iron than for protons, one expects  steady sources of heavy nuclei primaries to be comfortably allowed as UHECR producers.
 
Due to severe energy losses via photo-hadronic interactions with the cosmic photons, above $\sim60\,$EeV, only protons and iron-like heavy elements can survive propagation over distances larger than $\sim 50\,$Mpc (see, e.g., Fig.~3 of \citealp{KO11}). Intermediate mass primary nuclei can   reach the Earth only if produced very nearby  (for carbon-nitrogen-oxygen nuclei, 90\% should come from distances $\le 40$ Mpc, and 50\%  from $\le 20$ Mpc. For helium nuclei, almost 100\% should come from $\le 12$ Mpc).  Considering that the mass distribution is highly structured in the very nearby universe,   tighter bounds on the source number density are expected for intermediate nuclei, leading to a similar exclusion in spite of the relaxed luminosity bound.

Hence, the highest-energy cosmic rays are either iron-like heavy nuclei produced in steady sources, or generated in transient sources. 

It is currently difficult to discriminate between the  remaining three scenarios. Anisotropy studies with increased statistics from next-generation UHECR observatories should be able to constrain steady source populations, even for heavy nuclei composition \citep{2014A&A...567A..81R,Oikonomou15}. Anisotropy signatures expected from transient source scenarios are less straightforward to interpret than for steady candidates due to the time delay caused by magnetic deflections. 
Many studies can be found on the subject \citep{Murase_Takami09,Kalli10} but the ultimate probe of transient candidates will likely be a multi-messenger transient signal.\\

{\it Acknowledgements.}
We thank Kohta Murase, Aur\'elien Benoit-L\'evy, St\'ephane Charlot, Guilhem Lavaux,  Cole Miller, Richard Mushotzky, Paolo Privitera, and Joe Silk for helpful comments.    K.F. acknowledges the support of a Joint Space-Science Institute prize postdoctoral fellowship.
K.K. acknowledges support from the PER-SU fellowship at Sorbonne Universit\'es and from the Labex ILP (reference ANR-10-LABX-63, ANR-11-IDEX-0004-02). K.K. thanks the KICP and the University of Chicago for its kind support and hospitality.

\end{document}